\documentclass{revtex4}
\usepackage{amsmath}
\usepackage{amssymb}
\usepackage{graphicx}
\begin{document}

\title{5D Kaluza-Klein gravity: singularuty and freezing of $5^{th}$ dimension}
\author{Vladimir Dzhunushaliev
\footnote{Senior Associate of the Abdus Salam ICTP}} 
\email{dzhun@hotmail.kg} \affiliation{Dept. Phys. and Microel. 
Engineer., Kyrgyz-Russian Slavic University, Bishkek, Kievskaya Str. 
44, 720021, Kyrgyz Republic}
\author{Ratbay Myrzakulov }

\email{cnlpmyra@satsun.sci.kz} \affiliation{Institute of Physics and 
Technology, 480082, Almaty-82, Kazakhstan}


\begin{abstract}
Some reasonings are presented that the problem of a singularity in general 
relativity with the problem of freezing of the $5^{th}$ dimension can be 
connected. It is shown that some solutions in the 5D Kaluza-Klein gravity 
with the cross section in the Planck region have a region $(\approx
l_{Pl})$ where the metric signature changes from $\{ +,-,-,-,- \}$
to $\{ -,-,-,-,+ \}$. The idea is discussed that such switching can not 
occur following some general point of view that the Planck length is
the minimal length in the nature and consequently the physical
quantities can not change very quickly in the course of this length. 
In this case the dynamic of the $G_{55}$ metric component should be 
frozen. 
\end{abstract}

\maketitle

\section{Introduction}

One of the principal problems in physics is the problem of a singularity. 
This problem exists both in classical and quantum physics. In classical physics 
it is, for example, an infinite energy of a point-like charge in electrodynamics 
or singularities in the Schwarzschield and Reissner-Nordstr\"om solutions 
in general relativity. 
In quantum physics it is well known singularities connected with the loops of Feynman 
diagrams. The origin of these singularities is that the corresponding 
particles are point like. In classical electrodynamics it result in 
the fact that near to the point-like particle the electric field tends to 
infinity by such a manner that full energy is infinite. In general relativity 
exist general theorems which tell us that in some general sense the 
singularities are unavoidable. In this sense general relativity can not 
solve the problem of an infinite energy of point-like particle : 
Reissner-Nordstr\"om solution has a singularity at $r=0$. 
\par 
The modern string theory approach smooths this problem since the string 
is an extended object. Roughly speaking one can say that near to string the 
fields are not so singular in comparison with point-like particle. 
But string point of view has another disagreeable peculiarity : 
remaining within the framework of this theory we can not ask a question on an 
internal structure of string. In string theory the string is a fundamental 
structureless object and the question on an internal structure is the nonsensical 
question.
\par 
Thus we can think that the problem of the point-like singularity is connected 
with an internal structure of elementary particles. Probably the resolution 
of this problem can be received on the framework of a quantum theory of 
Everything including gravity. The problem here is that this quantum theory 
can not be based on the Feynman diagram technique since this approach from 
the very beginning on default assumes that quantum fields consist of 
point-like \textit{free} particles, \textit{i.e.} it is necessary to have 
a non-perturbative quantization method for strongly interacting fields 
(especially for gravity). Unfortunately now we do not have such method.
\par 
In this notice we would like to present some qualitative reasonings and 
quantitative calculations that the singularity problem can be closely connected 
with the problem of freezing (splitting off) extra dimensions. The idea is that 
all over the space the metric on the extra dimensions is not variable in the 
sense that we have not any equations for it. But near to a point-like 
singularity the gravitational field becomes so strong that excites the 
additional degrees of freedom : the metric on the extra dimensions. Such 
idea is not new but ordinary it is supposed that such compactification of 
the extra dimensions is a dynamical process which can be described on 
the language of field equations (classical or quantum, more probably that quantum). 
The idea presented here is that this process can not be described on the 
field equations level. The matter is that we suppose that on the level of Planck 
sizes fluctuate not only fields but field equations also. 
\par 
In this connection we would like to show that one can build a 
wormhole-like object with a very long 5-dimensional throat (and with the cross section 
in Planck region) and two asymptotically flat tails. The main point of view 
is that the wormhole is 5-dimensional object but on the tails $G_{55}$ 
metric component is frozen in the sense that the corresponding equation is absent 
but on the throat this equation is switched on. The main question is why 
this equation is switched on on the throat and switched off on the tails. 
Here we would like to show that such switching can not occur following some general 
point of view that the Planck length is
the minimal length in the nature and consequently the physical
quantities can not change very quickly in the course of this length. 
\par 
We will use this statement for the
metric signature. For example, the next dynamic is impossible: the
metric signature by $r < r_H - l_{Pl}$ is $\{ +,-,-,-,- \}$ and by
$r > r_H + l_{Pl}$ is $\{ -,-,-,-,+ \}$ ($r_H$ is some constant).
According to the above-mentioned statement in the region $|r - r_H|
< l_{Pl}$ have to be some quantum gravitational effects which will
prevent such dynamic. In this notice we do not give any
microscopical description how it happens but we consider only the
consequences of such prohibition. Such approach is similar to the
\emph{macroscopical} thermodynamical investigation of a physical
process when we do not interesting for the \emph{microscopical}
description of the process. 

\section{Wormhole-like flux tubes}

The 5-dimensional wormhole-like flux tube metric is
\begin{eqnarray}
    ds^2 & = & \frac{dt^{2}}{\Delta(r)} - \Delta(r) e^{2\psi (r)}
    \left [d\chi +  \omega (r)dt + Q \cos \theta d\varphi \right ]^2
    \nonumber \\
    &-& dr^{2} - a(r)(d\theta ^{2} +
    \sin ^{2}\theta  d\varphi ^2),
\label{sec1-10}
\end{eqnarray}
where $\chi $ is the 5$^{th}$ extra coordinate; $r,\theta ,\varphi$
are $3D$ spherical-polar coordinates; $r \in \{ - \infty, + \infty \}$
is the longitudinal coordinate; $Q$ is the magnetic charge.
\par
According the correspondence between 5-dimensional gravity and 4-dimensional gravity + electrodynamics + 
scalar field the metric \eqref{sec1-10} has the following components of the 
4-dimensional electromagnetic potential
$A_\mu$
\begin{equation}
A_t = \omega (r) \quad
\text{and} \quad
A_\varphi = Q \cos \theta .
\label{sec1-20}
\end{equation}
For this potential the Maxwell tensor is 
\begin{equation}
    F_{rt} = \omega' (r) \quad
    \text{and} \quad
    F_{\theta \varphi} = -Q \sin \theta .
\label{sec1-30}
\end{equation}
This means that we have radial Kaluza-Klein
electric $E_r \propto F_{tr}$ and magnetic
$H_r \propto F_{\theta \varphi}$ fields.
\par
Substituting this ansatz into the 5-dimensional Einstein vacuum equations
\begin{equation}
    R_{AB} - \frac{1}{2} \eta_{AB} R = 0
\label{sec1-35}
\end{equation}
$A,B = 0,1,2,3,5$ and $\eta_{AB}$ is the metric signature,
gives us
\begin{eqnarray}
    \frac{\Delta ''}{\Delta} - \frac{{\Delta '}^2}{\Delta^2} +
    \frac{\Delta 'a'}{\Delta a} + \frac{\Delta ' \psi '}{\Delta} +
    \frac{q^2}{a^2 \Delta ^2}e^{-4 \psi} & = & 0,
\label{sec1-40}\\
    \frac{a''}{a} + \frac{a'\psi '}{a} - \frac{2}{a} +
    \frac{Q^2}{a^2} \Delta e^{2\psi} & = & 0,
\label{sec1-50}\\
    \psi '' + {\psi '}^2 + \frac{a'\psi '}{a} -
    \frac{Q^2}{2a^2} \Delta e^{2\psi} & = & 0,
    \label{sec1-60}\\
    - \frac{{\Delta '}^2}{\Delta^2} + \frac{{a'}^2}{a^2} -
    2 \frac{\Delta ' \psi '}{\Delta} - \frac{4}{a} +
    4 \frac{a' \psi '}{a} +
    \frac{q^2}{a^2 \Delta ^2} e^{-4 \psi} +
    \frac{Q^2}{a^2} \Delta e^{2\psi} & = & 0
\label{sec1-70}
\end{eqnarray}
$q$ is the electric charge. These equations are derived after substitution
the expression
\eqref{sec1-80} for the electric field in the initial Einstein's equations 
\eqref{sec1-35}. The 5-dimensional $\binom{\chi}{t}$-Einstein's equation (4-dimensional Maxwell
equation) is taken as having the following solution
\begin{equation}
  \omega ' = \frac{q}{a \Delta ^2} e^{-3 \psi} .
\label{sec1-80}
\end{equation}
For the determination of the physical sense of the constant $q$ let us
write the $\binom{\chi}{t}$-Einstein's equation (from the 4-dimensional viewpoint it is the Maxwell 
equations) in the following way :
\begin{equation}
    \left( \omega ' \Delta ^2 e^{3\psi} 4 \pi a \right)' = 0.
\label{sec1-90}
\end{equation}
After the dimensional reduction of the 5-dimensional Kaluza - Klein gravity 
the Maxwell's tensor is
\begin{equation}
    F_{\mu \nu} = \partial_\mu A_\nu - \partial _\nu A_\mu .
\label{sec1-100}
\end{equation}
That allows us to write the electric field as $E_r = \omega '$.
Eq.\eqref{sec1-90} with the electric field defined by \eqref{sec1-100}
can be compared with the Maxwell's equations in a continuous medium
\begin{equation}
    \mathrm{div} \mathcal {\vec D} = 0
\label{sec1-110}
\end{equation}
where $\mathcal {\vec D} = \epsilon \vec E$ is an electric displacement
and $\epsilon$ is a dielectric permeability. Comparing Eq. \eqref{sec1-110}
with Eq. \eqref{sec1-90} we see that the magnitude
$q/a = \omega ' \Delta^2 e^{3\psi}$ is like to the electric displacement
and the dielectric permeability is $\epsilon = \Delta^2 e^{3\psi}$.
It means that $q$ can be taken as the Kaluza-Klein electric charge
because the flux of electric field is 
\begin{equation}
    \mathbf{\Phi} = 4\pi a\mathcal D = 4\pi q .
\label{sec1-110a}
\end{equation}
As the electric $q$ and magnetic $Q$ charges are
varied it was found \cite{dzhsin1} that the solutions to the metric in
Eq. \eqref{sec1-40}-\eqref{sec1-70} evolve in the following way :
\begin{enumerate}
\item
$0 \leq Q < q$. The solution is \emph{a regular gravitational flux tube}.
The solution is filled with both electric and magnetic fields. The longitudinal
distance between the $\pm r_H$ surfaces increases, and
the cross-sectional size does not increase as rapidly
as $r \rightarrow r_H$ with $q \rightarrow Q$. The values $r= \pm r_H$ are defined by the
following way $\Delta(\pm r_H)=0$.
\item
$Q = q$. In this case the solution is \emph{an infinite flux tube} filled
with constant electric and magnetic fields. The cross-sectional
size of this solution is constant ($ a= const.$).
\item
$0 \leq q < Q$. In this case we have
\emph{a singular gravitational gravitational flux tube}
located between two (+) and (-) electric and magnetic
charges at $\pm r_{sing}$. At
$r = \pm r_{sing}$ this solution has real singularities where are the charges.
\end{enumerate}
We will consider the case with $q \approx Q$ but $q > Q$. In this case
there is a region $|r| \leq r_H$ where the solution is like to
a tube filled with almost equal electric and magnetic fields.
The length $L = 2 r_H$ of the throat of the flux tube (by $|r| < r_H$) depends on relation
$\delta = 1 - Q/q$ by the following manner 
$L \stackrel{\delta \rightarrow 0}{\longrightarrow} \infty$
but $\delta > 0$. The numerical analysis \cite{dzh3} shows that the
spatial cross section of the tube ($t, \chi , r = \mathrm{const}$) does not change
significantly, for example, $a(r_H) \approx 2 a(0)$. The cross section of the tube
at the center $a(0)$ is arbitrary and we choose it as
$a(0) \approx l^2_{Pl}$. This allows us to say that we have a super-thin  and super-long
gravitational flux tube: $L \gg \sqrt{a(0)}$, $a(0) \approx l_{Pl}$.

\begin{figure}[h]
  \begin{center}
    \fbox{
    \includegraphics[height=7cm,width=7cm]{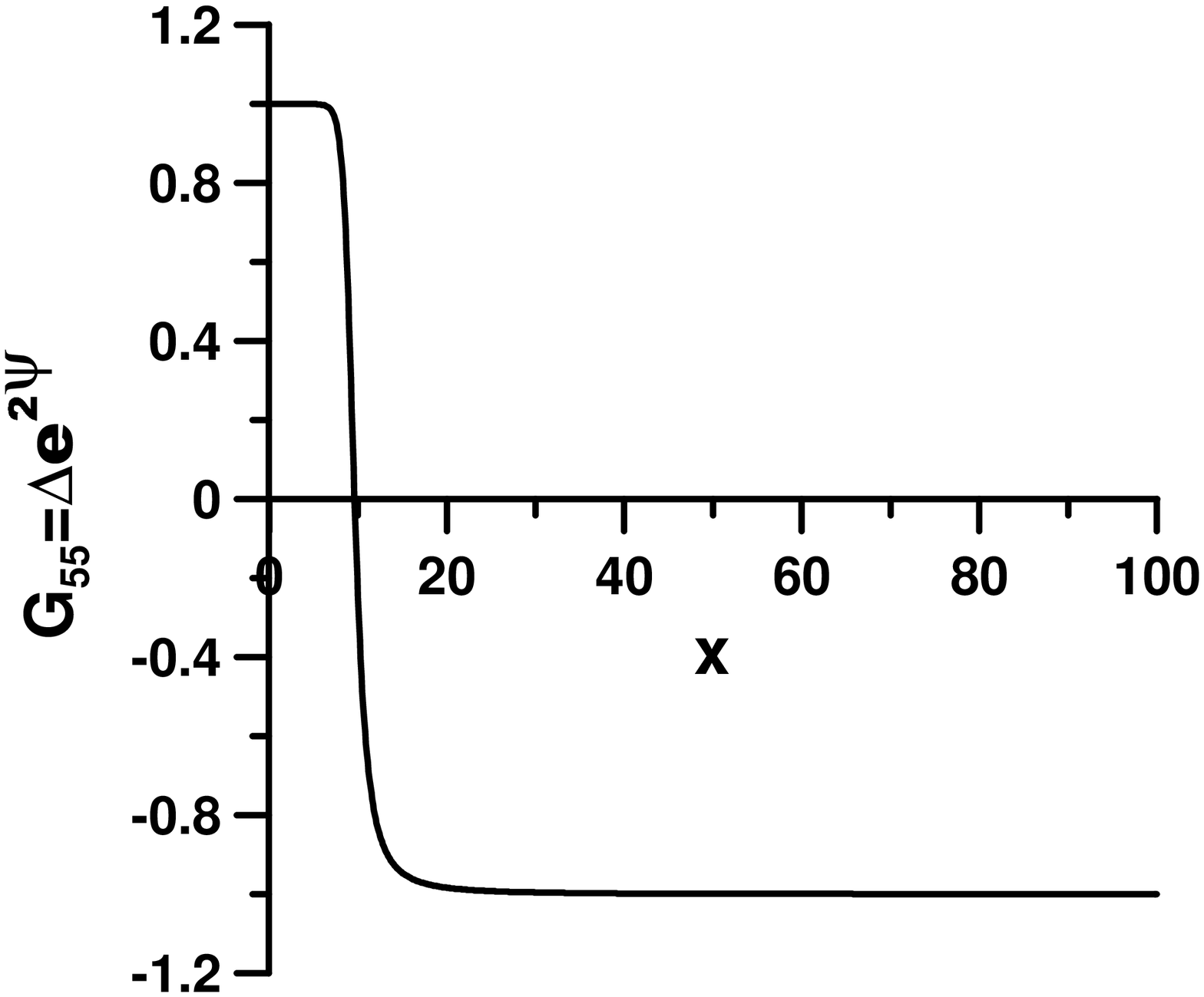}}
    \caption{The metric component $G_{55} = \Delta(x) e^{2\psi(x)}$, 
    $x = r/\sqrt{a(0)}$ is the dimensionless radius.}
    \label{fig:prod}
  \end{center}
\end{figure}
\par
On Fig. \eqref{fig:prod} the profile of the function
$\Delta \mathrm e^{2\psi}$ (which is equal to $G_{55}$)
from Ref. \cite{dzh3} is presented.
We see that nearby the value $r = r_H$ this function
changes drastically from the value $G_{55} \approx +1$ by
$r \gtrsim + r_H - l_0$ to $G_{55} \approx -1$ by $r \lesssim + r_H + l_0$. 
The same occur near $r = - r_H$. 
\par 
In order to present the metric on the throat by 
$\left| r \right| < r_H - l_0$ we present the exact solution with $q = Q$ 
\begin{eqnarray}
    a &=& a(0) = \frac{Q_0^2}{2} = const,
\label{sec1-130}\\
    e^{2\psi} &=& \frac{1}{\Delta} = \cosh^2 \frac{r}{\sqrt{a(0)}},
\label{sec1-140}\\
    \omega &=& \sqrt{2}\sinh\frac{r}{\sqrt{a(0)}} ,
\label{sec1-150}\\
        G_{55} & = & \Delta \mathrm e^{2 \psi} = 1
\label{sec1-155}
\end{eqnarray}
here we have parallel electric $E$ and magnetic $H$ fields with equal electric
$q_0$ and magnetic $Q_0$ charges $q_0 = Q_0 = \sqrt{2a(0)}$.
On the throat by $\left| r \right| < r_H - l_0$
(the solution with $q \approx Q$ but $q > Q$)
all equals signs of Eq's \eqref{sec1-130}-\eqref{sec1-155}
should be changed on the approximate equality signs
\begin{eqnarray}
    a & \approx & \frac{Q_0^2}{2} = const,
\label{sec1-160}\\
    e^{2\psi} & \approx & \frac{1}{\Delta} = \cosh\frac{r}{\sqrt{a(0)}},
\label{sec1-170}\\
    \omega & \approx & \sqrt{2}\sinh\frac{r}{\sqrt{a(0)}} ,
\label{sec1-180}\\
    G_{55} & = & \Delta \mathrm e^{2 \psi} \approx 1
\label{sec1-190}
\end{eqnarray}
Such approximation is valid only by $|r| \lesssim r_H - l_0$ where $l_0$ 
is some small quantity $l_0 \ll r_H$. 
\par
Now we would like to estimate the length $l_0$ of region
where the change of the metric component  $G_{55} = \Delta \mathrm e^{2\psi}$ is 
\begin{equation}
    \left. \Delta \mathrm e^{2\psi}\right|_{r \approx r_H + l_0} -
    \left. \Delta \mathrm e^{2\psi}\right|_{r \approx r_H - l_0} \approx 2 .
\label{sec1-195}
\end{equation}
For this estimation will be used Eq. \eqref{sec1-50}.
On the throat by $\left| r \right| < r_H - l_0$ this equation approximately is
\begin{equation}
    - \frac{2}{a} + \frac{Q^2}{a^2} \Delta e^{2\psi} \approx 0.
\label{sec1-200}
\end{equation}
We can estimate $l_0$ by solving Einsten's
equations \eqref{sec1-40}-\eqref{sec1-70} nearby $r = + r_H$
(for $r = - r_H$ the analysis is the same) and define $r = r_H - l_0$ where
the last two terms in Eq. \eqref{sec1-50} will have the same order
\begin{equation}
    \left.
        \left(
            \frac{2}{ a}
        \right)
    \right|_{r=r_H - l_0} \approx
    \left.
        \left(
            \frac{Q^2}{a^2} \Delta e^{2\psi}
        \right)
    \right|_{r=r_H - l_0}
\label{sec1-210}
\end{equation}
The solution close to $r = + r_H$ we search in the following form
\begin{equation}
    \Delta(r) = \Delta_1 \left( r_H - r \right) +
    \Delta_2 \left( r_H - r \right)^2 + \cdots .
\label{sec1-230}
\end{equation}
The substitution in Eq's \eqref{sec1-40} gives us the following
solution
\begin{equation}
    \Delta_1 = \frac{q e^{-2\psi_H}}{a_H}.
\label{sec1-260}
\end{equation}
After the substitution into Eq. \eqref{sec1-210} we have
\begin{equation}
    l_0 \approx \sqrt{a(0)} = l_{Pl}
\label{sec1-270}
\end{equation}
here we took into account that the numerical analysis \cite{dzh3}
shows that $a_H \approx 2 a(0)$.
It means that the change of macroscopical dimensionless function
$G_{55} = \Delta \mathrm e^{2\psi}$ as in Eq. \eqref{sec1-195} occurs
during the Planck length. The metric \eqref{sec1-10} by
$|r| \approx r_H - l_{Pl}$ approximately is
\begin{equation}
    ds^2 \approx \mathrm e^{2\psi_H} dt^2 - dr^2 - a(r_H)
    \left( d\theta^2 + \sin \theta d \varphi^2 \right) -
    \left(
        d\chi + \omega dt + Q \cos \theta d \phi
    \right)^2
\label{sec1-273}
\end{equation}
by $|r| \approx r_H + l_{Pl}$ the metric
approximately is
\begin{equation}
    ds^2 \approx -\mathrm e^{2\psi_H} dt^2 - dr^2 - a(r_H)
    \left( d\theta^2 + \sin \theta d \varphi^2 \right) +
    \left(
        d\chi + \omega dt + Q \cos \theta d \phi
    \right)^2
\label{sec1-276}
\end{equation}
here we took into account that numerical calculations \cite{dzh3} show
that $\psi \approx \psi_H = \mathrm {const}$ by $|r| \gtrsim r_H$.
We see that during the Planck length the metric signature changes
from $\{ +,-,-,-,- \}$ to $\{ -,-,-,-,+ \}$. Simultaneously it is necessary to
mention that the metric \eqref{sec1-10} is non-singular by $|r| = r_H$ and
approximately is \cite{dzh3}
\begin{equation}
\begin{split}
    ds^2 \approx &
    e^{2 \psi_H} dt^2 -
    e^{\psi_H} dt \left( d\chi + Q \cos\theta d \varphi \right) -
    dr^2 - a(r_H) \left( d\theta^2 + \sin^2 \theta d\varphi^2 \right) = \\
    & \left[
        e^{\psi_H} dt - \frac{1}{2} \left( d \chi + Q \cos \theta d \varphi \right)
    \right]^2 - 
    dr^2 - a(r_H) \left( d\theta^2 + \sin^2 \theta d\varphi^2 \right) - 
    \frac{1}{4} \left( d \chi + Q \cos \theta d \varphi \right)^2 
\label{sec1-280}
\end{split}
\end{equation}
where $\psi_H$ is some constant.
\par
If we write the metric \eqref{sec1-10} in the 5-bein formalism
\begin{equation}
\begin{split}
    ds^2 &= \omega^A \omega^B \eta_{AB} ,\\
    \omega^A &= e^A_\mu dx^\mu , \quad
    x^\mu = t,r,\theta , \varphi , \chi
\label{sec1-290}
\end{split}
\end{equation}
then we see that
\begin{eqnarray}
  \eta_{AB} &=& \left\{ +1,-1,-1,-1,-1 \right\} \quad
  \text{by} \quad |r| \lesssim r_H - l_{Pl}
\label{sec1-300}\\
  \eta_{AB} &=& \left\{ -1,-1,-1,-1,+1 \right\} \quad
  \text{by} \quad |r| \gtrsim r_H + l_{Pl}
\label{sec1-310}
\end{eqnarray}
It is necessary to note that for the mechanism presented here the change
of the sign of two components $\eta_{00}$ and $\eta_{55}$ is very important.
The reason is that the $G_{55}$ metric component can be made dimensional by such
a manner
\begin{equation}
    G_{55} d\chi ^2 = \left( l^2_0 G_{55} \right)\left( \frac{d \chi}{l_0} \right)^2
\label{sec1-320}
\end{equation}
where $l_0 \approx l_{Pl}$ is the characteristic length of the $5^{th}$
dimension. In this case the quantity $\sqrt{l_{Pl} G_{55}}$ changes
\begin{equation}
    \left. l_{Pl} \sqrt{G_{55}} \right|_{r \approx r_H + l_{Pl}} -
    \left. l_{Pl} \sqrt{G_{55}} \right|_{r \approx r_H - l_{Pl}}
    \approx l_{Pl}
\label{sec1-330}
\end{equation}
by changing the radial coordinate
\begin{equation}
    \Delta r \approx l_{Pl}.
\label{sec1-340}
\end{equation}
Such variation of $G_{55}$ is possible but simultaneously the dimensionless
quantity $\eta_{00} (e^0_t)^2$ changes
\begin{equation}
    \left. \eta_{00} (e^0_t)^2 \right|_{r \approx r_H + l_{Pl}} -
    \left. \eta_{00} (e^0_t)^2 \right|_{r \approx r_H - l_{Pl}}
    \approx 2 \mathrm e^{2\psi_H} \gg 1
\label{sec1-350}
\end{equation}
during the Planck length.
\par
But this is not the whole of history.

\subsection{Soft singularity}

Let us consider more carefully what occurs at the place $r = \pm r_H$ 
with the $G_{55}$ metric component. For this more convenient to present the metric 
\eqref{sec1-10} in the form 
\begin{equation}
\begin{split}
    ds^2 = \frac{e^{2\psi (r)}}{\tilde{\Delta}(r)} dt^2  - \tilde{\Delta}(r) 
    \left [d\chi +  \omega (r)dt + Q \cos \theta d\varphi \right ]^2 
    - dr^{2} - a(r)(d\theta ^{2} +
    \sin ^{2}\theta  d\varphi ^2),
\label{sec1-360}
\end{split}
\end{equation}
comparing the metrics \eqref{sec1-10} and \eqref{sec1-360} we see the following 
connection between functions $\Delta(r)$ and $\tilde{\Delta}(r)$ 
\begin{equation}
    G_{55} = \tilde{\Delta}(r) = \Delta(r) e^{2\psi (r)} .
\label{sec1-370}
\end{equation}
By definition the derivative of the $G_{55}$ is equal 
\begin{equation}
    \left.\tilde{\Delta}'(r) \right|_{r = r_H} = 
    \lim \limits_{\Delta r \rightarrow 0} 
    \frac{\tilde{\Delta} \left( r_H + \Delta r \right) - 
    \tilde{\Delta} \left( r_H - \Delta r \right)}{2 \Delta r} .
\label{sec1-380}
\end{equation}
Taking into account quantum gravity one can see that here there is one subtlety. 
The point is that $\Delta r$ can not converge to zero since there is the 
minimal length : Planck length. Ordinary it is not important but in our case 
we have a \textit{soft} singularity in the following sense. Equations 
\eqref{sec1-195} and \eqref{sec1-380} give us 
\begin{equation}
    \left.\tilde{\Delta}'(r) \right|_{r = r_H} \approx 
    \frac{\tilde{\Delta} \left( r_H + l_{Pl} \right) - 
    \tilde{\Delta} \left( r_H - l_{Pl} \right)}{2 l_{Pl}} = 
    \frac{\Theta \left( r_H + l_{Pl} \right) - 
    \Theta \left( r_H - l_{Pl} \right)}{2 l_{Pl}} \approx 
    \left.\Theta'(r) \right|_{r = r_H} 
    \approx \delta \left( r_H \right) = \infty
\label{sec1-390}
\end{equation}
where 
\begin{equation}
    \Theta \left( x \right) = 
    \left\{ 
    \begin{array}   {rl}
        +1, & \mbox{if } x > 0 \\
        -1, & \mbox{if } x < 0 
    \end{array} 
    \right.
\label{sec1-400}
\end{equation}
is the step function, $\delta(r)$ is the Dirac's delta function. It means that 
near to $r = \pm r_H$ 
\begin{equation}
    \tilde{\Delta}'(r) \approx \delta \left( r \right).
\label{sec1-410}
\end{equation}
It is not very bad if everywhere we would have only $\tilde{\Delta}'(r)$ 
not $\tilde \Delta '^2 (r)$. But it is not the case. For example the Ricci 
scalar is 
\begin{equation}
\begin{split}
    \frac{1}{2} R = &{\psi'}^{2}(r) - 
    \frac {\psi'(r) {\tilde{\Delta}'}(r)}{\Delta (r)} + 
    \frac {1}{2} \frac {\tilde \Delta'^2(r)}{{\tilde{\Delta}}^{2} (r)}  + 
    2 \psi'' (r) - 
    \frac {1}{2} \frac {\tilde{\Delta}^2 (r) \omega^2 (r)}{e^{2 \psi (r)}} + 
\\
    &2 \frac {a'(r) \psi' (r)}{a(r)}  - 
    \frac {1}{2} \frac {{a'}^2 (r)}{a^2 (r)}  + 
    2 \frac {a'' (r)}{a(r)}  - 
    \frac {2}{a (r)}  + 
    \frac {1}{2} \frac {Q^2 \tilde{\Delta} (r)}{a^2 (r)}.
\end{split}
\label{sec1-420}
\end{equation}
We see that we have $\tilde \Delta '^2 (r)$. It means that by $r = \pm r_H$ 
we have a singularity - \textit{a soft} singularity. The word soft signifies 
that in this case singularity differs from (\textit{hard}) singularity, 
for example, by $r = 0$ in the Schwarzschield black hole. 

\subsection{Pure quantum freezing of the $5^{th}$ dimension}

From the previous sections we saw that the gravitational flux tube solutions 
in the 5-dimensional Kaluza-Klein gravity have two regions where 
$G_{55}$ metric component just has such fast change. 
One of the basic paradigm of quantum gravity tell us that the Planck
length is the minimal length in the nature and consequently not any
physical fields and quantities can change in the course of the
Planck length. Consequently we can draw a conclusion that such classical dynamic
of the metric signature is \emph{impossible}. We suppose that only one way 
is to avoid such dynamical behavior of 
this quantity: \emph{any} dynamic of $G_{55}$ field variable should be 
forbidden with the conservation of its last value. One can say that 
it is \emph{a pure quantum freezing of the dynamic of $5^{th}$
dimension}. Mathematically it means that $G_{55}$ becomes non-dynamical
quantity and should not be varied by the derivation of Einstein's equations from
the corresponding Lagrangian.
\par
Thus we can say that the analysis of the classical dynamic of
the metric signature for the super-thin  and super-long gravitational
flux tube shows that there
is a region where this quantity changes too much quickly from the quantum
gravity viewpoint. It leads to the fact that some pure quantum gravity
effects have to happen in order to avoid such variation of
the metric signature. We do not
consider the mechanism of these effects but the authors point of view
is that such mechanism can not be based on any field-theoretical consideration.
This pure quantum freezing of extra dimension is like to a trigger which
has only two states: in one state the dynamic of $G_{55}$ is switched
on in another one is switched off. One can say that such quantum dynamic
which can be realized only in the Planck region is a non-differentiable
dynamic. The examples of such hypothesized non-differentiable dynamic
can be above-mentioned freezing of extra dimensions, the change of metric
signature and may be other phenomena \cite{Bogdanoff:2001dz}, \cite{castro}.
\par
The next question arising in this context is the dynamic of residuary
degrees of freedom. Since $G_{55} = \mathrm{const}$ we have Kaluza-Klein
gravity in its initial interpretation when 5-dimensional Kaluza-Klein theory with
$G_{55} = \mathrm{const}$ is equivalent to 4-dimensional electrogravity. It prompts an
idea that the spacetime with ($r \gtrsim r_H + l_{Pl}$)
(the same for $r \lesssim -r_H - l_{Pl}$) will be the Reissner-Nordstr\"om
solution with the corresponding electric and magnetic fields.

\section{Continuity of electric and magnetic fields}

In this section we would like to discuss the problems concerning to joining
the electric and magnetic fields of the gravitational flux tube 
considered above and the extreme Reissner-Nordstr\"om
solution. For this we will compare the fluxes of electric and magnetic
fields on the throat and the Reissner-Nordstr\"om spacetime.

\subsection{Continuity of electric field}

At first we consider the electric field. The Maxwell equations for the 
electric field on the curve space are
\begin{equation}
    \frac{1}{\sqrt{\gamma}} \frac{\partial}{\partial x^\alpha}
    \left(
        \sqrt{\gamma} \stackrel{(4)}{D}\!\!{}^\alpha
    \right) = 0
\label{sec2-10}
\end{equation}
where $\sqrt{\gamma}$ is the determinant of the 3D spatial metric and
$\stackrel{(4)}{D}\!\!{}^\alpha$ is an analog of the electric displacement in the
electrodynamics of continua medium in the presence of the gravitational field 
\begin{equation}
    \stackrel{(4)}{D}\!\!{}^\alpha = - \sqrt{g_{00}} \stackrel{(4)}{F}\!\!{}^{0 \alpha} .
\label{sec2-20}
\end{equation}
For the Reissner-Nordstr\"om spacetime the corresponding metric is
\begin{equation}
\begin{split}
    \stackrel{(4)}{ds}\!\!{}^2 = &\left( 1 - \frac{r_g}{R} + \frac{r_{q,Q}}{R^2} \right)
    \stackrel{(4)}{dt}\!\!{}^2 +
    \frac{dR^2}{1 - \frac{r_g}{R} + \frac{r_{q,Q}}{R^2}} -
    R^2 \left( d \theta^2 + \cos^2 \theta d \varphi^2 \right),\\
    & r_g = \frac{2 G m}{c^2}, \quad
    r_{q,Q} = \sqrt{\frac{G}{c^4} \left( q^2 + Q^2 \right)}
\end{split}
\label{sec2-30}
\end{equation}
where $G$ is the gravitational constant; $\stackrel{(4)}{(\:)}$ means that the 
corresponding quantity is 4-dimensional one; $R$ is the 4-dimensional radial 
coordinate. Therefore Eq. \eqref{sec2-10} is
\begin{equation}
    \frac{d}{d R}
    \left(
        R^2 \stackrel{(4)}{F}\!\!\!{}_{tR}
    \right) = 0 , \quad
    \stackrel{(4)}{F}\!\!{}_{tR} = \frac{d \phi}{dR}
\label{sec2-40}
\end{equation}
here $\phi$ is the scalar potential. This equation shows that the flux
$\stackrel{(4)}{\Phi}_e$ of the electric field $\stackrel{(4)}{E}_R = \phi '$ does 
conserve 
\begin{equation}
    \stackrel{(4)}{\Phi}\!\!{}_{e} = 4 \pi R^2 \phi ' = 4 \pi \!\!\stackrel{(4)}{q} =
    \mathrm{const} .
\label{sec2-50}
\end{equation}
The corresponding $\binom{5}{1}$ Einstein equation on the throat of the 5-dimensional 
flux tube is
\begin{equation}
     \frac{d}{d r}
     \left[
        a \left(
                \omega ' \Delta^2 \mathrm{e}^{3 \psi}
            \right)
     \right] = 0.
\label{sec2-60}
\end{equation}
In the consequence of Eq. \eqref{sec2-60} we have also the conserved flux of 
an analog of electric displacement
\begin{equation}
    \stackrel{(5)}{\Phi}\!\!{}_{e} = 4 \pi a \!\!\stackrel{(5)}{D} \!\!{}^r =
    4 \pi \!\!\stackrel{(5)}{q} = \mathrm{const}
\label{sec2-70}
\end{equation}
where
\begin{equation}
    \stackrel{(5)}{D} \!\!{}^r = \omega ' \Delta^2 \mathrm{e}^{3 \psi} =
    \omega ' \mathrm{e}^{-\psi} G_{55}.
\label{sec2-73}
\end{equation}
The simplest assumption about two electric fields on the throat and
Reissner-Nordstr\"om spacetimes is to join the fluxes \eqref{sec2-50} and
\eqref{sec2-70}
\begin{equation}
    \stackrel{(5)}{\Phi}\!\!{}_{e} = \stackrel{(4)}{\Phi}\!\!{}_{e}
\label{sec2-80}
\end{equation}
here $\stackrel{(5)}{\Phi}\!\!\!{}_{e}$ have to be calculated at the ends of throat where
freezing of $5^{th}$ coordinate happens, i.e. by $r \approx r_H - l_{Pl}$ 
on the right side of the throat and $r \approx -r_H + l_{Pl}$ 
on the left side of the throat. There $G_{55} \approx 1$ and
\begin{equation}
    \phi ' \approx \omega '\mathrm{e}^{-\psi_H}
\label{sec2-90}
\end{equation}
that results to
\begin{equation}
    \stackrel{(5)}{q} = \stackrel{(4)}{q} = q
\label{sec2-100}
\end{equation}
here we have used the conditions
\begin{eqnarray}
    \left. \stackrel{(4)}{g}\!\!\!{}_{\theta \theta} \right|_{R = r_0} & = & r_0^2 =
    a(0) = \left. \stackrel{(5)}{G}\!\!{}_{\theta \theta} \right|_{r = r_H - l_{Pl}}
\label{sec2-110}\\
    \psi_H & \approx & \left. \psi \right|_{r = r_H - l_{Pl}} = 
    \left. \psi \right|_{r = - r_H + l_{Pl}}
\label{sec2-120}
\end{eqnarray}

\subsection{Continuity of magnetic field}

Now we will repeat the similar calculations for the magnetic field. The Maxwell
equation for the magnetic field having the information about the flux of magnetic
field is the same for the throat and the tails (= Reissner-Nordstr\"om spacetimes)
\begin{equation}
    \frac{1}{\sqrt{\gamma}} \frac{\partial}{\partial x^\alpha}
    \left[
        \sqrt{\gamma} \left(
            \frac{-1}{2\sqrt{\gamma}} \epsilon^{\alpha\beta\delta} F_{\beta\delta}
        \right)
    \right] = 0
\label{sec2b-10}
\end{equation}
here we introduce the magnetic field $B^\alpha$ 
\begin{equation}
    B^\alpha = \frac{-1}{2\sqrt{\gamma}} \epsilon^{\alpha\beta\delta} F_{\beta\delta}.
\label{sec2b-20}
\end{equation}
More concretely Eq. \eqref{sec2b-20} has the following form for the 
Reissner-Nordstr\"om spacetime 
\begin{equation}
    \frac{d}{dR} \left[
        R^2 \left(
            \frac{\stackrel{(4)}{Q}}{R^2}
        \right)
    \right] = 0
\label{sec2b-30}
\end{equation}
and similar for the 5-dimensional case 
\begin{equation}
    \frac{d}{dr} \left[
        r^2 \left(
            \frac{\stackrel{(5)}{Q}}{r^2}
        \right)
    \right] = 0
\label{sec2b-35}
\end{equation}
here $r, R$ are the radial coordinates on the throat and the tails correspondingly. 
It allows us to
introduce the flux of magnetic field for the throat and the tails which will be equal
\begin{equation}
    \stackrel{(4)}{\Phi}\!\!{}_m = 4 \pi R^2 \frac{\stackrel{(4)}{Q}}{R^2} =
    4 \pi a \frac{\stackrel{(5)}{Q}}{a} = \stackrel{(5)}{\Phi}\!\!{}_m .
\label{sec2b-40}
\end{equation}
Therefore one can introduce correctly the magnetic charge
\begin{equation}
    \stackrel{(4)}{Q} = \stackrel{(5)}{Q} = Q .
\label{sec2b-50}
\end{equation}

\subsection{Continuity of metric}

After quantum freezing of the $5^{th}$ dimension 5-dimensional metric 
\eqref{sec1-10} will be
\begin{equation}
    \stackrel{(5)}{ds}\!\!{}^2 = A(r)dt^{2} - B(r) dr^{2} - a(r)(d\theta ^{2} +
    \sin ^{2}\theta  d\varphi ^2) -
    \left [d\chi +  \omega (r)dt + Q \cos \theta d\varphi \right ]^2 .
\label{sec2c-10}
\end{equation}
The interpretation of this metric in the Kaluza-Klein gravity tells us
that we have 4-dimensional metric
\begin{equation}
    \stackrel{(4)}{ds}\!\!{}^2 = A(r)dt^{2} - B(r) dr^{2} - a(r)(d\theta ^{2} +
    \sin ^{2}\theta  d\varphi ^2)
\label{sec2c-20}
\end{equation}
and the electromagnetic potential
\begin{equation}
    A_\mu = \left\{ \omega ,0,0, Q \cos \theta \right\}.
\label{sec2c-30}
\end{equation}
In the previous subsections we have joined the electric and magnetic fields.
Now we have to consider the metric components. We will interpret the metric
\eqref{sec2c-20} as Reissner-Nordstr\"om metric. After freezing of $5^{th}$
coordinate the corresponding equations become 5-dimensional Einstein's equations but
without $\binom{5}{5}$ Einstein equation, i.e. we have the ordinary 
4-dimensional electro-gravity.
Taking this into account the metric \eqref{sec2c-10} will be Reissner-Nordstr\"om
metric \eqref{sec2-30}. We will join $g_{\theta\theta}$ and $g_{tt}$ components
of the metrics \eqref{sec1-10} and \eqref{sec2c-20}. It is necessary to note that joining
the $g_{rr}$ components of these metrics is not necessary as they measure
the distance on the transversal direction to the surface of joining.
We do not join (as usually) the first derivatives of metric components since
on the 5-dimensional throat and the tails there are the different equations set: 
on the throat - 5-dimensional Einstein's equations but on the tails 4-dimensional 
electro-gravity equations : let us remember that according to the initial interpretation 
of the Kaluza-Klein gravity the 4-dimensional electro-gravity spacetime can be considered 
as 5-dimensional spacetime with frozen $5^{th}$ dimension ($G_{55} = \mathrm{const}$).
\par
Joining of $g_{\theta \theta}$ gives
\begin{equation}
    \stackrel{(4)}{g}\!\!{}_{rr} \left( r_0 \right) =
    \left. a \right|_{\pm r_H \mp l_{Pl}} \quad
    \Rightarrow \quad r_0^2 = a(0) = l^2_{Pl}.
\label{sec2c-40}
\end{equation}
For the $g_{tt}$ components
\begin{equation}
    \left(
        1 - \frac{r_g}{r_0} + \frac{r^2_{q,Q}}{r_0^2}
    \right)
    \stackrel{(4)}{dt}\!\!{}^2 = \mathrm{e}^{2\psi_H}\stackrel{(5)}{dt}\!\!{}^2
\label{sec2c-50}
\end{equation}
here $\stackrel{(5)}{t}$ and $\stackrel{(4)}{t}$ are the time coordinates
on the 5-dimensional throat and the 4-dimensional tails correspondingly. 
One can rewrite this relation as
\begin{equation}
    \frac{\stackrel{(5)}{dt}}{\stackrel{(4)}{dt}} =
    \mathrm{e}^{-\psi_H}
    \sqrt{1 - \frac{r_g}{l_{Pl}} + \frac{r^2_{q,Q}}{l_{Pl}^2}}.
\label{sec2c-60}
\end{equation}
Only by such relation between the time coordinates on the 5-dimensional throat 
and the 4-dimensional tails the time will pass equally on the hypersurface 
of joining.
\par
It is necessary to mention that the analysis presented in this section is very
simple and, for example, do not give us the possibility to determine the mass
$m$ for the Reissner-Nordstr\"om solution. Probably more exact calculations
could be made on the quantum field-theoretical language.

\section{Conclusions and discussion}

Thus in this notice we wanted to show that in quantum gravity there are 
some reasonings speaking about that some classical dynamic of $G_{55}$ 
is forbidden. We have shown that the super-thin  and super-long
gravitational flux tube solutions in 5-dimensional Kaluza-Klein gravity have
two regions where the classical description can not be applied. Some
metric components change too quickly: the metric signature changes
from $\eta_{ab} = \{ +,-,-,-,- \}$ to $\eta_{ab} = \{ -,-,-,-,+ \}$ 
and consequently $\Delta G_{55}
\approx 2$ during the Planck length. To avoid such variation the dynamic 
of the metric component $G_{55}$ must be frozen. Then the metric on the throat 
can be extended via the right and left ends to the
Reissner-Nordstr\"om spacetimes. If we additionally will freeze 4-dimensional 
metric $g_{\mu \nu}$ then we will have only Maxwell 
electrodynamic on the flat space which is similar to the AdS/S
correspondence idea. At the joint places takes place \emph{a pure
quantum freezing of the dynamic of $G_{55}$ metric component}. Such
object looks as two extremal Reissner-Nordstr\"om spacetimes
($r^2_g/4 < r^2_{q,Q}$) connected with a super-thin  and super-long
flux tube filled with the electric and magnetic fields. Let us note
that the presented here point of view is based only on the statement
which does not depend on the details of one or another quantum
gravity theory.
\par
Such model allow us successfully to resolve one of the hardest problems of
general relativity: the avoidance of a singularity. The construction 
presented here shows that for an extremal Reissner-Nordstr\"om solution 
(at least for $q > Q$) the gravitational field near to singularity 
becomes so strong that the dynamic on the $5^{th}$ dimension becomes excited
and a flux tube from one singularity to another one appears. The force lines do not
convergence in a pointlike singularity but leave our universe to another one
(or to a remote part of our universe) through the flux tube and there appear
near to another singularity. Similar idea in the 4-dimensional case 
is presented in Ref. \cite{zaslavski} as a model of an elementary particle. 
\par
There is another argument for presented here the pure quantum freezing mechanism.
We see that by $|r| \lesssim r_H - l_{Pl}$ the metric component $G_{55} \approx 1$ 
is almost frozen. It means that with big accuracy we have \emph{dynamical} 
freezing of the $5^{th}$ coordinate on the throat but near to the surface of the change 
of metric signature the quantity $G_{55}$ becomes dynamical. One can 
suppose that there is a quantum mechanism which keep this freezing from dynamical 
one to pure quantum one.
\par 
It is necessary to note that such mechanism does work only for some special extremal
Reissner-Nordstr\"om solutions having the electric and magnetic charges with
the relation $q > Q$. Certainly the question arises: is it possible to extend
the result presented here about avoidance of a singularity to other 
Reissner-Nordstr\"om solutions ?
In this connection we have to emphasize that the analysis carried out in tis notice 
is very simple and more careful analysis with the quantization of
corresponding field quantities can give more exact results. In this connection one can
mention the results of Ref. \cite{Dzhunushaliev:2004wj} where it is shown that the super-thin 
and super-long flux tube solutions without freezing $5^{th}$ dimension have interesting
properties on the tails: the magnetic fields decreases faster the electric
field and there is a rotation connected with the magnetic field on the throat.
One can presuppose that these properties will remain the same by switching on
freezing $5^{th}$ coordinate (may be in some weaker form).
\par
The construction presented here (two Reissner-Nordstr\"om spacetimes 
connected by a flux tube) can be considered as a model of a 
spacetime with frozen $5^{th}$ dimension but with piecewise 
impregnations where the $5^{th}$ dimension is unfrozen. The similar 
idea about spacetime regions with compactified and uncompactified 
phases was considered by Guendelman in Ref. 
\cite{Guendelman:1991pc}. The construction presented in 
\cite{Guendelman:1991pc} is also based on the flux tube (Levi-Civita 
- Bertotti - Robison solution \cite{levi-civita}-\cite{robinson}) 
inserted between two 4-dimensional black holes. Similarly in Ref. 
\cite{zaslavski} the model of classical electron with Levi-Civita - 
Bertotti - Robison flux tube between two Reissner-Nordstr\"om black 
holes is considered.

\end{document}